\documentclass[12pt]{elsart}
\usepackage{graphics}
\usepackage{epsfig}
\usepackage{setspace}
\usepackage{longtable}

\begin{document}
\begin{frontmatter}
\title{Lubricated friction between incommensurate substrates}
\author{A. Vanossi$^a$\corauthref{cor}},
\ead{vanossi.andrea@unimore.it} \corauth[cor]{Corresponding
Author.}
\author{G. E. Santoro$^{b,c}$},
\author{N. Manini$^{d}$},
\author{E. Tosatti$^{b,c}$},
and \author{O. M. Braun$^{e}$}
\address{$^a$CNR-INFM National Research Center S3 and
Department of Physics, \\University of Modena and Reggio Emilia,
Via Campi 213/A, 41100 Modena, Italy}
\address{$^b$International School for Advanced Studies (SISSA)
and INFM Democritos National Simulation Center, Via Beirut 2,
I-34014 Trieste, Italy}
\address{$^c$International Centre for
Theoretical Physics (ICTP), P.O.Box 586, I-34014 Trieste, Italy}
\address{$^d$Department of Physics, University of Milan, Via Celoria 16, 20133 Milan, Italy}
\address{$^e$Institute of Physics, National Academy of Sciences of Ukraine, 03028 Kiev, Ukraine}

\maketitle
\begin{abstract}
This paper is part of a study of the frictional dynamics of a
confined solid lubricant film -- modelled as a one-dimensional
chain of interacting particles confined between two ideally
incommensurate substrates, one of which is driven relative to the
other through an attached spring moving at constant velocity. This
model system is characterized by {\it three} inherent length
scales; depending on the precise choice of incommensurability
among them it displays a strikingly different tribological
behavior. Contrary to {\it two} length-scale systems such as the
standard Frenkel-Kontorova (FK) model, for large chain
stiffness one finds that here the
most favorable (lowest friction) sliding regime is achieved by
chain-substrate incommensurabilities belonging to the class of
non-quadratic irrational numbers (e.g., the spiral mean). The
well-known golden mean (quadratic) incommensurability which slides
best in the standard FK model shows instead higher
kinetic-friction values.
The underlying reason lies in the pinning properties of
the lattice of solitons formed by the chain with the substrate
having the closest periodicity, with the other slider.
\end{abstract}

\end{frontmatter}

\section{Introduction}

Nonlinear systems driven far from equilibrium exhibit a very rich
variety of complex spatial and temporal behavior. In particular,
in the emerging field of nanoscale science and technology,
understanding the nonequilibrium dynamics of systems with many
degrees of freedom which are pinned in some external potential, as
commonly occurs in solid-state physics, is becoming more and more
a central issue. Classic textbook examples are the motion of
dislocations in metals, of domain walls in ferroelectrics, the
problem of crowdion in a metal, sub-monolayer films of atoms on
crystal surfaces, transport in Josephson junctions, the sliding of
charge density waves and so on. Sliding friction of solids belongs
to this class of systems as well, because the microscopic periodic
asperities of the mating surfaces may interlock.

One of the pervasive concepts of modern tribology with a wide area
of relevant practical applications as well as fundamental
theoretical issues is the idea of free sliding connected with {\it
incommensurability}. When two crystalline workpieces with lattices
that are incommensurate (or commensurate but not perfectly
aligned) are brought into contact, the minimal force required to
achieve sliding (i.e., the {\it static friction}, $F_s$) should
vanish, provided the two substrates are stiff enough. In this
configuration, the lattice mismatch causes the total energy to be
completely independent of the relative position of the sliders,
whereas hardness can prevent pinning and the associated stick-slip
motion of the interface atoms, with a consequent negligibly small
frictional force. The remarkable conclusion of frictionless
sliding can be drawn, in particular, in the well explored 1D
context of the Frenkel-Kontorova (FK) model (see \cite{BK1998} and
references therein). Experimental observation of this kind of {\it
superlubric} regime has recently been reported \cite{graphite}.

Based on the success of this kind of simplified tribological
approach \cite{Vanossi_review}, the hope is to be able to learn
more from the direct study of microscopic dynamics in similarly
simplified 1D models. The substrates defining the sliding
interface are modelled as purely rigid surfaces or as 1D (or 2D)
arrays of particles interacting through simple (e.g., harmonic)
potentials. Despite its extreme idealization, this kind approach
seems able to anticipate features of the actual experimental
results or of more complex molecular dynamics simulations of
frictional phenomena. In this context, the application of {\it
driven} FK like models (see \cite{BK2004} and references therein),
describing the dissipative dynamics of a chain of interacting
particles that slide over a rigid periodic substrate potential due
to application of an external driving force, has found an
increasing interest as a possible interpretative key to understand
the atomic processes occurring at the interface of two materials
in relative motion. The essential feature of the static and
dynamic properties of the FK model consists in the competition
between the interparticle interaction (having natural lattice
constant $b$) and the substrate periodic potential (of spatial
period $a$). Indeed, if the former favors a uniform separation $b$
between particles, the latter tends to pin the atom positions to
the bottom of the wells, evenly spaced by the period $a$. This
length-scale competition, often referred to as frustration,
results in a fascinating complexity of spatially modulated
structures. For this {\it two} competing length-scale system,
following the pioneering work of Aubry \cite{Aubry}, the
tribological properties depend mainly on the  value of the winding
number $a/b$ being rational (commensurate interface) or irrational
(incommensurate interface). Rational interfaces are always
ingrained and frictional, whereas irrational interfaces exhibit an
``Aubry transition'' from frictional to frictionless as the chain
stiffness increases.

\begin{figure}[p]
\centerline{
\epsfig{file=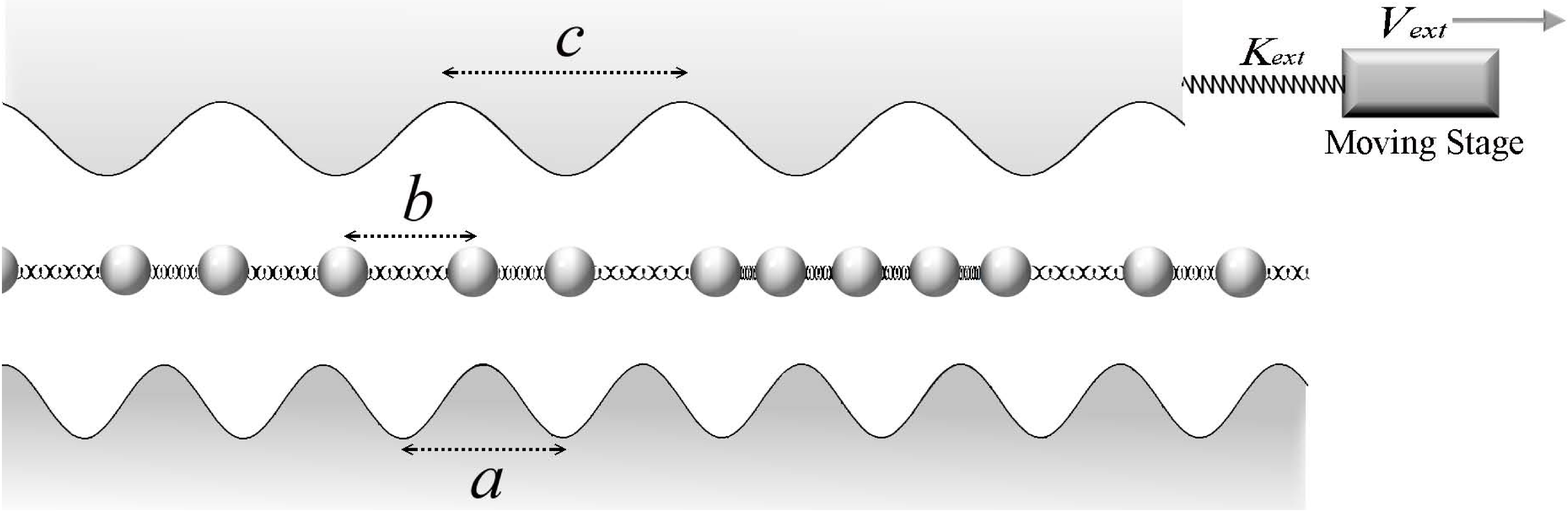,width=15cm,angle=0} }
\caption{\label{model} Schematic drawing of the lubricant chain
confined between the two rigid substrates. As shown, the sliding
interface is characterized by the three inherent length scales:
$a$, $b$ and $c$.}
\end{figure}

In real situations, however, such a case of ``dry'', unlubricated
friction is often exceptional. The physical contact between two
solids is generally mediated by so-called ``third bodies'', which
may act like a lubricant film. If the third body is a solid
lubricant, and in particular a crystalline one, the sliding interface
corresponds in fact to a system with {\it three} inherent lengths:
the periods of the bottom and top substrates, and the period of
the lubricant layer. As sketched in Fig.~\ref{model}, in order to
study the role of incommensurability among the three interface
inherent lengths on the sliding dynamics, we may consider a 1D
generalized FK model consisting of {\em two} rigid sinusoidal
substrates, of spatial periodicity $a$ (bottom) and $c$ (top),
harassing a chain of harmonically interacting particles, of
equilibrium length $b$, mimicking the confined lubricant. Guided
by a similarity with the recent study of the driven quasiperiodic
FK model \cite{vanossi}, we show that, for sufficiently stiff
chain (hard crystalline lubricant), the best low-friction regime
is achieved for incommensurabilities $a/b$ and $c/a$ related
through non-quadratic irrational numbers (as, for example, the
cubic spiral mean \cite{spgl}) rather than through quadratic
irrationals (e.g., the golden mean). This result demonstrates how,
in a {\it three}-length sliding system, the {\em kind} of
interface incommensurability (and not just the distinction between
commensurate and incommensurate geometry) can dramatically
influence, both quantitatively and qualitatively, the tribological
behavior.

\section{Model and kinetic friction}

In numerical simulations of frictional processes, different
approaches may be used to induce motion: a constant-velocity
algorithm, where the top substrate is directly forced to slide
with a given velocity, or a constant-force algorithm, where the
external driving force is applied to the top substrate (or even
directly to the lubricant layer), or finally a spring-force
algorithm, where the slider is driven through an attached spring
moving at constant speed. The latter pulling procedure can be
viewed as a way to mimic not only the experimental driving device,
but to some degree also the elasticity of the slider.

In this paper, we consider a 1D system of two rigid sinusoidal
substrates and a chain of interacting particles embedded between
them \cite{PRL}, as shown in Fig.~\ref{model}. The top substrate
of mass $M$ is driven through a spring, $K_{\rm ext}$, which is
pulled with constant velocity $V_{\rm ext}$. The equations of
motion become
\begin{eqnarray}
m \ddot{x}_i &+& \gamma \dot{x}_i + \gamma (\dot{x}_i-\dot{X}_{\rm
top}) + \frac{d}{dx_i} \sum_{i \ne j} V(|x_i-x_j|) \nonumber \\
&& \hspace{7mm} +\frac{1}{2} \left[ \sin{\frac{2 \pi x_i}{a} } +
\sin{\frac{2 \pi (x_i-X_{\rm top})}{c}} \right] = 0,
\label{MOTION1}
\\
M \ddot{X}_{\rm top} &+& \sum\limits_{i = 1}^N \gamma
(\dot{X}_{\rm top}-\dot{x}_i) + K_{\rm ext}(X_{\rm top}-V_{\rm
ext}t) \nonumber \\
&& \hspace{7mm} + \sum\limits_{i = 1}^N \frac{1}{2} \left[
\sin{\frac{2 \pi (X_{\rm top}-x_i)}{c}} \right] = 0, \hspace{5mm}
\label{MOTION2}
\end{eqnarray}
where $x_i$ ($i=1,\ldots,N$) and $X_{\rm top}$ stand for the
coordinates of the $N$ chain particles and the top substrate,
respectively. The damping $\gamma$-terms describe the dissipative
forces which are proportional to the relative velocities of the
lubricant atoms with respect to both rigid substrates. The viscous
coefficient $\gamma$ represents degrees of freedom inherent in the
real physical system (such as substrate phonons, electronic
excitations, etc.) which are not explicitly included in the model.
The last (sinusoidal) terms represent the on-site interaction
between the particles and the substrates. The amplitudes of these
rigid potentials are chosen such that the same factor ($1/2$) sits
in front of their derivatives in the equations of motion -- whence
the periodic forces they produce have the same magnitude. The
interparticle chain interaction [fourth term in
Eq.~(\ref{MOTION1})] is harmonic with strength $K$ and equilibrium
spacing $b$. We use dimensionless units with chain atom mass $m=1$
and a bottom substrate period $a=1$. We focus here on the
tribologically most interesting case of the {\it underdamped}
regime, where the damping coefficient $\gamma$ is much smaller
than both characteristic vibrational frequencies of one particle
sitting at the minima of the two substrate potentials.

The system is initialized with the lubricant particles placed at
rest at uniform separation $b$. After relaxing the starting
configuration, the stage attached to the top substrate through the
spring $K_{\rm ext}$, begins to move at constant speed $V_{\rm
ext}$. The equations of motion (\ref{MOTION1}), (\ref{MOTION2}) are
integrated using the standard fourth-order Runge-Kutta algorithm.
After reaching the steady state, the relevant physical
characteristics of the system are measured. The detailed behavior
of the driven system in Eqs.~(\ref{MOTION1}), (\ref{MOTION2}) depends
crucially on the length-scale relative ratios $a/b$ and $c/a$ of
the substrates and the chain. Since commensurability can only be
achieved under well-controlled experimental operating conditions
\cite{Muser}, in this investigation we focus on a few specific but
typical cases of incommensurability among the three length scales
$a, b, c$. Specifically we assume a single ratio $a/b=c/a$ and
consider the two irrational cases previously studied in the
context of the driven quasiperiodic FK model \cite{vanossi},
namely the golden mean and the spiral mean \cite{spgl}.
Nevertheless, it should be noted that the qualitative features of
the observed dynamics (in a similar 1D confined model
corresponding to the $M\to \infty$ limit of Eq.~(\ref{MOTION2}),
so that in Eq.~(\ref{MOTION1}) $\dot{X}_{\rm top}\equiv V_{\rm
ext}$) have been shown recently to survive for much more general
values of $a/b$ and $c/a$ \cite{recent1,Santoro06,recent2}, so
that this specific choice of incommensurability should not be
considered too restrictive. By imposing periodic boundary
conditions in order to simulate an infinite system, we are forced
to approximate the desired irrational winding numbers $a/b$, $c/a$
by ratios of integers. Continued-fraction expansion
\cite{fraction} provides a standard and well-established technique
to serve this purpose. Quantitatively, a given rational
approximant can be checked to be of sufficiently high order to
model the desired incommensurability. The chain of $N$ atoms and
length $L=Nb$ is thus confined in between the bottom and top
substrates with $N_a=L/a$ and $N_c=L/c$ minima, respectively.

In tribology applications, one of the main issues is the kinetic
friction force $F_{\rm kin}$ which describes the energy losses
inside the contact layer when the substrates move relatively to
each other. In our model, $F_{\rm kin}$ can be easily evaluated
from energy balance. When the top substrate moves for a distance
$\Delta X_{\rm top}=\dot{X}_{\rm top}\Delta t$ relative to the
fixed bottom substrate, due to the friction force the system
dissipates an energy
\begin{equation}
E_{\rm loss}=F_{\rm kin}\, \Delta \! X_{\rm top}= F_{\rm kin}
\dot{X}_{\rm top} \, \Delta t. \label{Eloss1}
\end{equation}
In the regime of steady motion, this must balance the losses due
to the dissipation generated by the $\gamma$-damping terms in the
chain, i.e.,
\begin{equation}
E_{\rm loss}=\sum\limits_{i = 1}^N \int\limits_{t_1}^{t_2} dt \;
\gamma \left[ \dot{x}_i^2+(\dot{x}_i-\dot{X}_{\rm top})^2 \right],
\label{Eloss2}
\end{equation}
where $\Delta t=t_2-t_1$. Thus, for the kinetic friction force, we
obtain
\begin{equation}
F_{\rm kin}=\sum\limits_{i = 1}^N \lim \limits_{\Delta t
\rightarrow \infty} \int\limits_{t_1}^{t_2} \frac{dt}{\Delta t} \;
\gamma \left[ \dot{x}_i^2+(\dot{x}_i-\dot{X}_{\rm top})^2 \right]
\frac{1}{\dot{X}_{\rm top}}. \label{Fkin}
\end{equation}
In order to highlight in the kinetic friction force the
dissipative part resulting from the internal dynamics of the
confined chain, we subtract the systematic contribution
arising from the motion of the top substrate, and also normalize
$F_{\rm kin}$ to the number of particles $N$, i.e.,
\begin{equation}
f_{\rm kin}=\frac{F_{\rm kin}}{N}- \frac{1}{2} \gamma \dot{X}_{\rm
top}. \label{fknorm}
\end{equation}

\section{Results}

In the range of model parameters considered, our numerical
simulations reveal a strikingly different sliding behavior of the
system depending on incommensurability. In particular, it is not
sufficient simply to distinguish between commensurate and
incommensurate periodicities. The ``degree'' of incommensurability
(as measured by continued-fraction expansion) plays a fundamental
role as well. Here the mathematical properties of the length-scale
ratio leave the realm of abstract number theory to become
physically relevant. In the context of the $M\to\infty$ confined
model, the peculiar incommensurability-dependent phenomenology has
been recently explained \cite{recent1} by the top slider (meaning
here that one among the two substrates whose period $c$ is the
longest, so that $b < a < c$) rigidly ``dragging'' the topological
solitons, with linear density $ \rho_s = 1/b - 1/a$, that the
chain forms with the bottom substrate.

In particular, the choice of incommensurability among the three
length scales $a, b, c$ governs the sliding behavior by inducing a
{\em commensurate} or {\em incommensurate} matching of the soliton
lattice in the chain with respect to the top substrate. In the
golden-mean case, $\rho_s c = \phi^2 - \phi = 1$, the soliton
lattice is commensurate with the top slider, and so it must
ingrain with it. In the spiral mean case, $ \rho_s c = \sigma^2 -
\sigma$ is irrational, the soliton lattice is incommensurate with
the top slider, and so it can slide freely over it. That describes
precisely what happens \cite{tobepubl}.

\begin{figure}
\centerline{
\epsfig{file=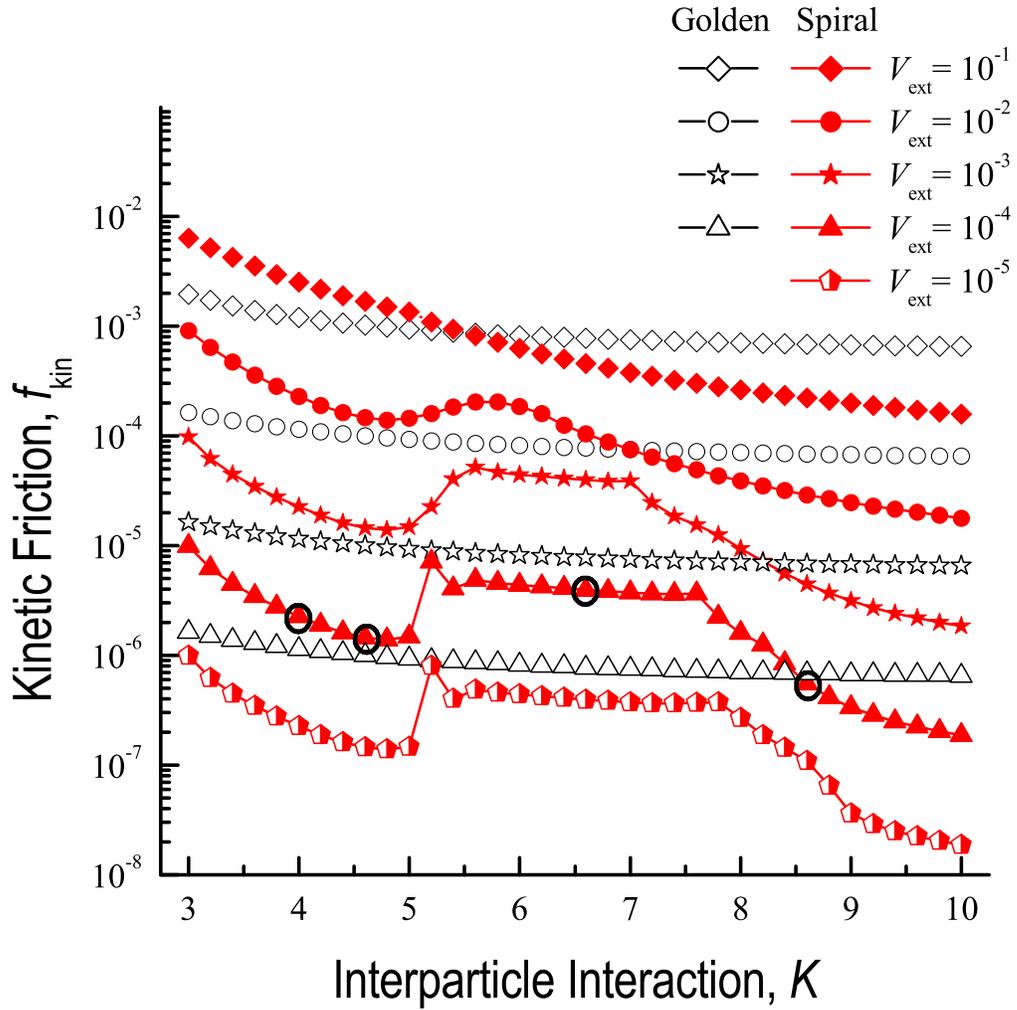,width=15cm,angle=0} }
\caption{\label{Fkinetic} Dependence of the kinetic friction force
$f_{\rm kin}$ on the chain stiffness $K$ for different values of
the external driving velocity. Both cases of golden and spiral
incommensurability are shown.}
\end{figure}

Figure~\ref{Fkinetic} shows the kinetic friction force $f_{\rm
kin}$ as a function of the interparticle interaction strength $K$
for different values of the velocity of the moving stage. The
plots are given for two irrational choices of the three length
scales: the quadratic golden mean ($N_a=144, N=233, N_c=89$) and
the cubic spiral mean ($N_a=265, N=351, N_c=200$). By simulating
larger systems, we checked that these rational approximants
\cite{spgl} are of sufficiently high order to mimic the selected
incommensurabilities.  In all simulations we set $M=N_c \, m$, and
take $\gamma=0.2$.

\begin{figure}
\centerline{
\epsfig{file=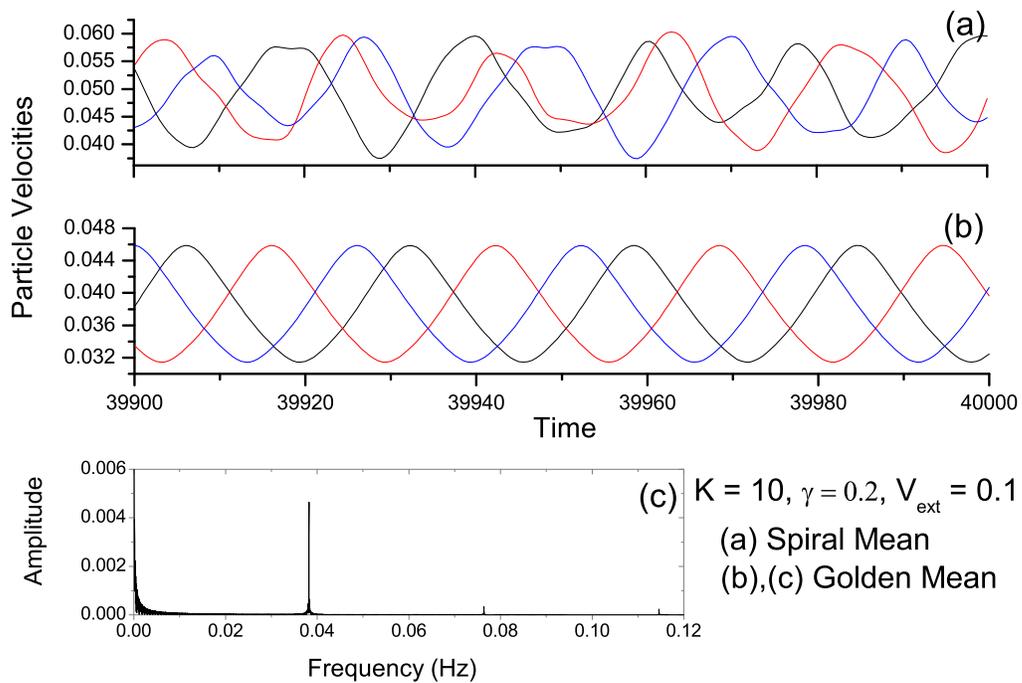,width=15cm,angle=0} }
\caption{\label{GoldSpiral} Atomic velocities versus time for
(a)~spiral incommensurability $a/b=351/265$, $c/a=265/200$;
(b)~golden incommensurability $a/b=233/144$, $c/a=144/89$. For
clarity, only the velocities of 3 adjacent particles of the chain
are shown. Panel (c) displays the corresponding Fourier
transform of $\dot x_i$, showing the sharp peak exactly located at the
frequency $\nu = 1/T \approx 0.03819$ that characterizes the
time-periodic dynamics of the golden-mean case.}
\end{figure}

For the \textit{golden\/} incommensurability, the qualitative
(monotonic) behavior of the kinetic friction does not depend
significantly on either $K$ or $V_{\rm ext}$. For the same
value of the chain stiffness, $f_{\rm kin}$ decreases with
decreasing driving velocity. The chain dynamics exhibits an
asymmetric sliding with respect to the two sinusoidal substrates,
moving as expected with an intermediate mean velocity, but always
faster with respect to the top substrate with the longer
spatial period. We checked that the asymmetry persists even when
the amplitudes of the substrate potentials are varied
significantly with respect to each other. This robust sliding
state characterized by a regular time-periodic dynamics, as
demonstrated in Fig.~\ref{GoldSpiral}(b), is specific to
the golden incommensurability and to all other quadratic
irrationals that we considered. Each single particle in the chain
performs exactly the same periodic motion, differing simply by a
phase shift. The period $T \approx 26.180$ of the motion is
highlighted by the corresponding frequency peak
$\nu = (1-\frac ba)\,V_{\rm ext} \simeq 0.038197$
in the Fourier transform of the velocity $\dot x_i$ of one
generic particle (Fig.~\ref{GoldSpiral}(c)).

As anticipated above, specific quadratic irrationals $\lambda$, such as the
golden mean, occur to satisfy $ \rho_s c = \lambda^2 - \lambda = m/n$, so
that the soliton lattice and the top slider are commensurate.
This commensurability provides an intrinsic ingraining which extends to
in principle arbitrarily large chain stiffness, without any transition.
As an additional rationale, one may imagine that this particularly strong
asymmetric center of mass motion is precisely such as to excite, via its
time-periodic feature, parametric resonances inside the chain
\cite{resonances}, thus converting part of the center-of-mass translational
kinetic energy into internal vibrational excitations, making in turn the
golden irrationality less favorable to sliding than the spiral one at the
values of $K$ considered here.

For inherent lengths related instead by the cubic
\textit{spiral\/} mean, the function $f_{\rm kin}(K)$ is found to
display a non-monotonic and richer behavior, with several
different regimes. Only at high driving velocities, $V_{\rm ext} >
0.01$, these exotic features diminish and eventually disappear.
The single-particle motion (see Fig.~\ref{GoldSpiral}(a))
of the chain, as confirmed by the
Fourier spectrum of $\dot x_i$ (not shown), is now definitely not
periodic. This suppresses the
parametric resonances inside the confined layer as possible
channels for energy dissipation, making, for stiff chains (large
$K$), the spiral-ratio slider more effective than the golden-ratio
one. As noted above, in the spiral case the soliton lattice is
incommensurate with the upper slider, and when stiff enough it may
actually slide essentially freely over it. The evolution at large
$K$ seems in fact towards an almost frictionless smooth sliding.
For smaller stiffness, a sharp change in the sliding
behavior of the spiral chain is observed at a critical stiffness
$K_c \simeq 5.2$, near the value where the Aubry (static depinning)
transition for the quasiperiodic FK model was found
previously \cite{vanossi}.
When the driving velocity decreases (see
lower curves in Fig.~\ref{Fkinetic}), a plateau region of {\em larger}
dynamic friction $f_{\rm kin}(K)$ develops, in sharp contrast to the
unpinning observed in the static behavior.
As can be seen, the location of its left edge is almost
independent of $V_{\rm ext}$ and centered around the value $K_c$.
For $K<K_c$ finally the dynamics is characterized by a stick-slip
motion. This is consistent with the forced sliding of an
incommensurate, but slack and therefore pinned, soliton irregular
arrangement. At large and increasing chain stiffness, the spiral
friction drops considerably (with a center of mass velocity
becoming symmetric, i.e. $V_{\rm ext}/2$), as compared to the
golden-mean case: in the latter, friction retains quite large
values due to the velocity asymmetry of its fully regular
time-periodic motion which is not destabilized until strikingly
large $K$.

\begin{figure}
\centerline{
\epsfig{file=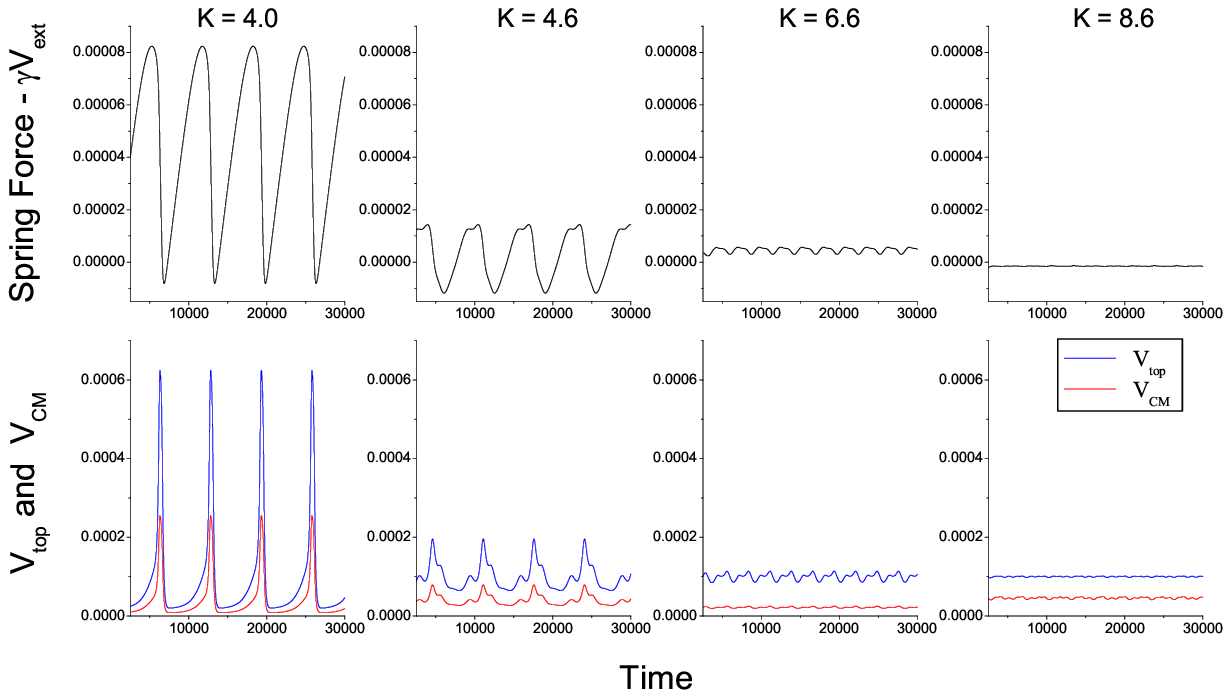,width=15cm,angle=0} }
\caption{\label{4states} Dynamics of the four sliding states
marked by the black circles in Fig.~\ref{Fkinetic}. The four upper
panels show the temporal behavior of the spring force, to which the trivial
$\gamma \, V_{\rm ext}$ drift term is subtracted, with the
characteristic transition from stick-slip motion at low $K$
(left) to smooth sliding at high $K$ (right). The lower panels
display the corresponding time dependencies of the velocities of
the top substrate and of the center of mass of the lubricant
chain.}
\end{figure}

In order to get more insight into different regions to the left,
in between, and to the right of the spiral frictional plateau, we
analyze in detail the four points circled in black in
Fig.~\ref{Fkinetic}, all belonging to the same driving velocity
$V_{\rm ext}=10^{-4}$. The four upper panels of Fig.~\ref{4states}
show the time evolution of the spring force $F_{\rm ext} \equiv
K_{\rm ext}(X_{\rm top}-V_{\rm ext}t)$; the lower panels display
the corresponding velocities of the top substrate and of the
center of mass of the chain. The first upper panel for $K=4.0$
clearly displays a typical sawtooth dependence of $F_{\rm
ext}(t)$, the hallmark of stick-slip motion. Stick-slip as a
phenomenon is quite general and extremely well known. It has been
observed in many simulations of microscopic friction, as well as
in the tip dynamics of atomic force microscopy and in surface
force apparatus experiments of confined lubricants under shear
(see, e.g., Ref.~\cite{Muser} and references therein). Energy is
stored in the springs while atoms are trapped in a metastable
state (sticking), and it is converted to kinetic energy as the
atoms pop to the next metastable state (slipping). The value of
the spatial distance covered during the slip events and the
regular-periodic or intermittent-chaotic temporal alternation
between stick and slip can be affected by several factors: the
intrinsic periodicity of the lattices defining the substrate
potentials, the potential energy landscape of accessible
metastable states, the specific features of imposing driving, etc.
At $K=4.6$ the slip events are prevailing, but the stick-slip
regime of the chain is still clearly visible.
The next value examined, $K=6.6$, lies in the plateau region. Here
the chain stiffness is too high to support stick-slip, and the
confined layer motion reaches a sliding regime with almost
constant value of the pulling spring force, and the velocity of
the top substrate exhibiting only very tiny oscillations around
$V_{\rm ext}$. The increase in the average kinetic friction
$f_{\rm kin}$ of the plateau states (relative to the contiguous
stick slip dynamics region) can be ascribed to the asymmetric
chain motion with respect to the two substrates in this regime. In
this range of stiffness and driving velocity values, a better
dynamical phase-matching is possible between the confined layer
and the two sinusoidal substrates, promoting the excitation of
dissipative parametric resonances, similar to the golden case.
Inside this plateau, even if the detailed particle dynamics still
remains irregular (quasiperiodic), a remarkable phase cancellation
between the Fourier spectra of different chain particles (all
having the same amplitude spectrum, with different phases) yields
strictly periodic center-of-mass motion of the embedded chain.
Finally, at $K=8.6$ the confined layer slides almost freely with a
very low value of $f_{\rm kin}$.
Sliding becomes symmetric in this
regime, i.e.\ the average chain velocity nearly equals one half of
the relative velocity of the two rigid substrates.

\section{Conclusions}

When two periodic substrates are forced to slide relative to each
other but a solid ``lubricant'' layer is interposed in
between, the frictional dynamics of the intermediate layer may
display an unexpected behavior, significantly dependent on the
incommensurability chosen. Our results show that, depending on physical
parameters such as the lubricant stiffness, at least for a
$3$-lengths contact interface (a typical physical situation) there
is no quantitative and qualitative uniformity of behavior in
incommensurate sliding friction, and that certain incommensurate
(e.g., non-quadratic) geometries can favor sliding systematically
better than others. In particular, we showed here that for
sufficiently stiff confined chains, the standard golden-mean case
reveals a higher kinetic friction than the spiral one. This is
attributable to the hidden commensurability of the soliton lattice
formed between the chain and the lower slider, which occurs in the
golden mean but not in the spiral mean case. It is not
clear at the moment to what extent this conclusion can be generalized
from the present 1D case to more realistic
2D interfaces. Similarly to what was recently done
for a real $2$-length sliding contact (graphite flake on graphite)
\cite{graphite}, our three length situation could, in principle,
be accessed experimentally by driving, e.g., a graphite
flake between two different but well defined crystalline surfaces.

\section*{Acknowledgments}
This research was partially supported by PRRIITT (Regione Emilia
Romagna), Net-Lab ``Surfaces \& Coatings for Advanced Mechanics
and Nanomechanics'' (SUP\&RMAN).


\begin{thebibliography}{99}
\bibitem{spgl} The spiral mean ($\sigma \approx 1.3247$) belongs
to the class of cubic irrationals, satisfing the equation
$\omega^3 - \omega -1=0$. Its rational approximants can be
generated by the recursion relation $G_{n+1}=G_{n-1}+G_{n-2}$ with
$G_{-2}=G_0=1, G_{-1}=0$. For the golden mean
$\phi=(1+\sqrt{5})/2$, the needed rational approximants are
provided instead by the famous Fibonacci sequence
$F_{n+1}=F_{n}+F_{n-1}$ with $F_0=F_1=1$.
\end{thebibliography}
\end{document}